\documentstyle[aasms4,12pt]{article}

\begin{document}

\title{A DYNAMICAL STUDY OF GALAXIES IN THE HICKSON COMPACT GROUPS}

\author{\sc Shingo Nishiura$^{1,2}$, Masashi Shimada$^{1,2,3}$,
        Youichi Ohyama$^{2,4}$, Takashi Murayama$^{1,2}$, and
        Yoshiaki Taniguchi$^{1,2}$}

\affil{$^1$Astronomical Institute, Graduate School of Science,
       Tohoku University, Aramaki, Aoba, Sendai 980-8578, Japan}
\affil{$^2$Visiting astronomer of Okayama Astrophysical Observatory, National
       Astronomical Observatory of Japan}
\affil{$^3$PENTAX}
\affil{$^4$National Astronomical Observatory of Japan, 2-21-1 Osawa, Mitaka,
           Tokyo 181-8588, Japan}

\begin{abstract}
In order to investigate dynamical properties of 
spiral galaxies in the Hickson compact groups (HCGs),
we present rotation curves of 30 galaxies in 20 HCGs. 
We found as follows. 
1) There is not significant relation 
between dynamical peculiarity and morphological peculiarity 
in HCG spirals. 
2) There is no significant relation 
between the dynamical properties and the frequency distribution of 
nuclear activities in HCG spirals. 
3) There are no significant correlations between 
the dynamical properties of HCG spirals and any group properties 
(i.e., the size, the velocity dispersion, the galaxy number density, 
and the crossing time). 
4) Asymmetric and peculiar rotation curves are more frequently 
seen in the HCG spirals than in field spirals and in cluster ones. 
However, this tendency is more obviously seen in late-type HCG spirals. 
These results suggest that the dynamical properties of HCG spirals do 
not strongly correlate with the morphology, the nuclear activity, 
and the group properties. 
Our results also suggest that more frequent galaxy collisions occur 
in the HCGs than in the field and in the clusters. 
\end{abstract}


\clearpage


\section{INTRODUCTION}

Dynamical properties of a galaxy are basically governed by both 
the mass and angular-momentum distributions in the galaxy.
Since such dynamical properties are deeply related to the
formation and evolution of galaxies, many dynamical studies
have been done for various kinds of galaxies (e.g., Rubin et al. 1985;
see for recent papers, Rubin, Waterman, \& Kenney 1999; Sofue
et al. 1999).
Dynamical properties also provide important 
information on the interaction between galaxies 
(Keel 1993, 1996; Chengalur et al. 1994; M\'arquez \& Moles 1996;
Barton, Bromley, \& Gellar 1999) and on the galaxy environment
such as clusters of galaxies (Rubin et al. 1988, 1999;
Whitmore et al. 1988). 

In addition to the central region of clusters of galaxies,
compact groups (CGs) of galaxies are also useful laboratories
to investigate violent interactions between/among
galaxies because they are small and isolated systems
whose galaxy number densities are comparable to those of the center
of cluster of galaxies; e.g., $\sim 10^{4-6}$ galaxies Mpc$^{-3}$
(Shakhbazyan 1973; Rose 1977; Hickson 1982). 
Among such CGs, the Hickson compact groups (HCGs) of galaxies
have been studied extensively (Hickson 1982, 1993).
Many galaxies in the HCGs show peculiar morphologies such as tidal tails,
tidal bridges, distorted isophotes, shell structures, and so on
(Mendes de Oliveira \& Hickson 1994). It is also known that 
a number of early-type galaxies in the HCGs have unusually blue colors
(Zepf et al. 1991; Moles et al. 1994). 
These observational results suggest that galaxies in the HCGs have 
experienced frequent dynamical interactions. 
Indeed, Rubin et al. (1991) showed that 
rotation curves of many spiral galaxies in HCGs appear abnormal.

In order to investigate the effect of dynamical interactions
in the CG environment, we newly conducted an optical spectroscopy
program of HCG galaxies (Shimada et al. 2000: Paper I). 
This paper presents results of statistical studies 
with rotation curve properties of HCG spirals. 
We describe our observation and the data reduction in section 2. 
Making rotation curve, estimation of rotation curve asymmetry, 
classifying rotation curve shapes are described in section 3. 
In section 4 we compare the rotation curve properties 
of the HCG spiral galaxies with 
the optical morphologies, nuclear activities, group properties 
(the group size, the velocity dispersion, the galaxy number density, 
and the crossing time).  
We compare the rotation curve properties between the HCG spirals 
and field ones in section 5 and clusters ones in section 6.  
In section 7 we discuss our results. 

We adopt a Hubble constant $H_{0} = 100$ km s$^{-1}$ Mpc$^{-1}$ and
a deceleration parameter $q_0 = 0$  throughout this paper.


\section{OBSERVATIONS}

We have obtained optical long-slit spectra along major-axis of 
30 galaxies (mostly disk galaxies) in 20 HCGs. 
The sample galaxies were selected  randomly 
from the HCG catalog (Hickson 1993).
The optical spectroscopy was made 
using the new Cassegrain spectrograph with an SITe 512 
$\times$ 512 CCD camera attached to the 188 cm telescope 
at the Okayama Astrophysical Observatory (OAO) 
during a period between 1996 February and 1997 January. 
A journal of the observations is given in Table 1. 
Basic data of the observed galaxies from Hickson (1993) are 
summarized in Table 2.
As for the morphology type, we preferentially adopted 
the Hubble Type taken from de Vaucouleurs et al. (1991, hereafter RC3). 
For galaxies whose Hubble type are 
uncertain in RC3, we adopted the Hubble type taken from Hickson (1993).

A long (5 arcmin) slit with a width of 1.8 arcsec was used and put on 
each target galaxy with a position angle of the major axis. 
The 600 grooves mm$^{-1}$ grating was used to cover 
6300 -- 7050 \AA ~ region with the spectral 
resolution of 3.4 \AA ~ ($\simeq$ 157 km s$^{-1}$ in velocity at 6500 \AA).
Two-pixel binning was made of the CCD along the slit and 
thus the spatial resolution was 1.75 arcsec per element. 
The typical seeing during the runs was 2 arcsec.
  
The data were analyzed using IRAF\footnote{%
Image Reduction and Analysis Facility (IRAF) is
distributed by the National Optical Astronomy Observatories,
which are operated by the Association of Universities for Research
in Astronomy, Inc., under cooperative agreement with the National
Science Foundation.}. We also used a special data reduction 
package, SNGRED (Kosugi et al. 1995), developed for OAO new Cassegrain 
spectrograph data. The reduction was made with a standard procedure; bias 
subtraction, flat fielding with the data of the dome flats, and cosmic ray 
removal. Flux calibration was obtained using standard stars available in IRAF. 
 

\section{RESULTS}

\subsection{Major-axis Velocity Curves}

We use the H$\alpha$ emission line to construct a heliocentric 
velocity curve as a function of the radial distance ($r$) from the nucleus
for each galaxy; i.e., $V_{\rm obs}(r) =
c [\lambda_{\rm obs}(r)/\lambda_{0} - 1]$ 
where $\lambda_{\rm obs}(r)$  and $\lambda_0$ are the measured and the
rest-frame wavelengths of H$\alpha$, respectively.
For each galaxy, we derive the rotation velocity curve correcting for
the inclination effect; i.e., 
$V(r) = [V_{\rm obs}(r)-V_{\odot}]/
[(1 + V_{\odot}/c) \sin i]$ where $V_{\odot}$ is the heliocentric velocity of 
the galaxy center, 
$c$ is the light velocity, and  $i$ is the inclination angle of 
the galaxy ($i = 0^{\circ}$ corresponds to the face-on view).
Following Rubin et al. (1982), 
we estimate the inclination angle using a relation of
$\sin i = 1.042^{0.5}(1 - 10^{-2x})^{0.5}$ where 
$x = \log (R_{\rm major}/ R_{\rm minor})$;
$R_{\rm major}$ and $R_{\rm minor}$ are the semimajor and 
semiminor axis of the isophote at 25 mag arcsec$^{-2}$ 
in the {\it B} band, respectively (Hickson 1993). 
The adopted values of $\sin i$ are 
given in Table 2. 
Distances from galactic nucleus $r$, rotation velocities $V(r)$ 
and their 1 $\sigma$ fitting error $dV(r)$ are given in Table 15 
(Appendix). 
In Figure 1, we show the rotation velocity curves as a function of 
distance from the galactic nucleus in units of arcsec 
for the observed galaxies. 

\subsection{Asymmetry of the Rotation Curve}

It is known that galaxy collisions disturb the rotation curves of galaxies.
The kinematical effect due to the tidal disturbance is often
different between the side facing to the colliding partner and
the opposite side and thus the rotation curve tends to show an 
asymmetrical property (Barton et al. 1999 and references therein).
Therefore, it is interesting to investigate the asymmetry of the
rotation curve. In order to quantify the asymmetry of the rotation curve,
we define an asymmetry parameter $A$, 

\begin{equation}
A \equiv \left[ \frac{1}{N}\sum^{N}_{j} \left( 
\frac{\left[V(r_{j})-V(-r_{j})\right]}{\left[V(r_{j})+V(-r_{j})\right]} 
\right)^{2} \right]^{1/2},  
\end{equation}
where $j$ is the bin number along the major axes and 
$N$ is the total bin number. Here we use the data between $r=0.2R_{25}$ 
and $r=0.5R_{25}$, where $R_{25}$ is length of the radius 
of the isophote at 25 mag arcsec$^{-2}$ in the $B$ band. 
The reason for this is as follows. The minimum radius, 
$0.2R_{25}$, is adopted to exclude the data in the central 
region of galaxies 
where the tidal disturbance is expected to be negligibly small; 
i.e., if we include the data with $r < 0.2R_{25}$, the difference of 
the asymmetry parameter among the galaxies could be less pronounced. 
On the other hand, the maximum radius, $0.5 R_{25}$, is adopted 
to cover the observed rotation curves for most of the galaxies 
studied here.
According to the definition of $A$, galaxies with higher asymmetric rotation
curves tend to have larger values of $A$. 
The results are given in Table 3. 
We cannot estimate $A$ for six spirals (HCG 37b, 47a, 61c, 87a, 92c, 
and 96a) because of the small data points in their rotation curves.

\subsection{Shape of the Rotation Curves}

As shown in Figure 1, the observed velocity curves show various shapes.
However, we simply adopt following three types of shapes; 
1) Type ``f'': the rotation velocity 
monotonically rises near the center and tend to be almost 
flat at $r/R_{25} \leq 1$; note that most ordinary spiral galaxies 
have this type of rotation curves (e.g., Rubin et al. 1985), 
2) Type ``fp'': the rotation curve 
appears almost flat but some dips and/or bumps are seen, 
3) Type ``p'': the rotation curve shows a significantly peculiar 
shape. Rotation curves with a sinusoidal shape or 
a linearly-rising shape are included in this type. 
The results of our classification are given in Table 3. 
Note that rotation curves of two galaxies 
(HCG 37b, and 92c) cannot be classified because 
there are only few data points.

\section{THE ROTATION CURVE PROPERTIES OF HCG GALAXIES}

\subsection{Enlarged HCG Sample}

Among the 30 galaxies observed by us, two galaxies are 
redshift-discordant galaxies in the HCGs (HCG 73a and 92a). 
One of remaining 28 galaxies is classified as an S0 
in RC3 (HCG 87a). Therefore, our HCG sample contains 
27 spiral galaxies. 
In order to enlarge the sample, we add HCG galaxies observed by
Rubin et al. (1991) (see Table 4). Their sample contains 34 spiral 
galaxies. Excluding two no-available galaxies (HCG 10a and 37c) 
of which rotation curve data were not shown, two redshift-discordant 
galaxies (HCG 31d and 78a), and five S0 galaxies classified in RC3 
(HCG 16c, 16d, 23a, 34b, and 57e), we obtained the rotation curve data 
of the 25 HCG spirals from Rubin et al. (1991). 
For these 25 HCG spiral galaxies, 
we estimate the asymmetry parameter and classify 
the rotation curve shape with the same method for our HCG data. 
The results are also given in Table 4.
Twelve of these 25 spiral galaxies are commonly observed 
(HCG 31a, 31b, 31c, 37b, 40c, 44a, 44b, 44d, 79d, 88a, 88c, and 89a). 
The $A$ values of these twelve spirals are adopted the averaged $A$. 
Finally, we obtain an enlarged HCG sample which contains 40 spiral galaxies. 
Hereafter, we discuss the rotation curves of HCG spirals with 
the enlarged HCG spiral sample containing 40 spirals. 

\subsection{Rotation Curve Properties versus Morphologies of Host Galaxies}

Mendes de Oliveira \& Hickson (1994) showed that about a half of 
galaxies in HCGs have peculiar morphologies such as tidal tails, 
tidal debris, distorted 
isophotes, and so on, indicating evidence for 
galaxy collisions. On the other hand, Rubin et al. (1991) found 
that there is a loose correlation between peculiarity of rotation curve 
and peculiarity of morphology. In Table 5, we classified our HCGs spiral 
galaxies into two categories based on the optical morphology 
from Mendes de Oliveira \& Hickson (1994) and RC3. 
The spiral galaxies with or without optical peculiar morphologies 
are listed in Table 5. 
We compare the shape of rotation curves with the optical morphologies 
of host galaxies. The result is shown in Table 6. 
Although it is generally expected that spiral galaxies with 
normal rotation curve have normal morphology and those 
with peculiar rotation curve have peculiar morphology, 
there are normal morphology galaxies with peculiar kinematics, 
and peculiar morphology galaxies with normal kinematics. 
We adopted the null hypothesis that distribution 
for the HCG spiral galaxies with peculiar morphology is the same as that 
for with normal morphology and apply the $\chi^{2}$ test. The result is 
shown in Table 6. We obtained a probability, $P(\chi^{2})=0.818$. 
We find that there is no significantly statistical 
difference between the HCGs spiral galaxies with peculiar morphology 
and that with normal morphology. 

\subsection{Rotation Curve Properties versus Nuclear Activities}

It has often been considered that 
frequent galaxy collisions trigger some nuclear 
activities (active galactic nuclei or intense star formation) 
in HCG member galaxies. 
Therefore, it is interesting to compare the kinematical properties 
in the member galaxies with the nuclear activity. 

Using observed flux ratio of $[{\rm N}{\sc ii}]$ to H$\alpha$, we 
classify the activities of galaxies into active galactic nuclei 
(AGN) and H{\sc ii} nucleus galaxies (H{\sc ii}); galaxies with 
$f([{\rm N}{\sc ii}])/f({\rm H}\alpha) \geq 0.6$ are classified into 
AGN, and those with $f([{\rm N}{\sc ii}])/f({\rm H}\alpha) < 0.6$ are 
classified into H{\sc ii} (Ho et al. 1997). 
The $f([{\rm N}{\sc ii}])/f({\rm H}\alpha$) ratio and 
the activity type for each galaxy are listed in Table 5. 
In Figure 2, we compare the distributions of the asymmetry parameter 
$A$ between AGNs and H{\sc ii} nuclei. We apply 
the Kolmogorov-Smirnov (KS) test (e.g. Press et al. 1988). 
The null hypothesis is that the distributions of $A$ of 
both the AGN and the H{\sc ii} nuclei come from the same underlying 
population. We obtain a KS probability, 0.947. This indicates that 
there is no difference in the distribution of the asymmetry parameter 
between the HCG AGN and the HCG H{\sc ii} nuclei.  
In Figure 3, we show diagrams of $[{\rm N}{\sc ii}]/{\rm H}\alpha$ 
ratio against the asymmetry parameter $A$. 
The left panel is the diagram for the case of early-type 
(S0/a-Sbc) and late-type (Sc and later) spirals 
and the right panel is the one for the case of ``f'' rotation curves, 
``fp'' ones, and ``p'' ones. 
We adopt the null hypothesis that the asymmetry parameter $A$ is not 
correlated with $[{\rm N}{\sc ii}]/{\rm H}\alpha$ ratio and 
apply the Spearman-rank statistical test for 
all the case shown in Figure 3. 
A summary of the statistical tests is given in Table 7. 
There is no correlation 
between the $A$ and $[{\rm N}{\sc ii}]/{\rm H}\alpha$ ratio 
in the HCG spirals.  

\subsection{Rotation Curve Properties versus Group Properties}

Rubin et al. (1991) mentioned that there appears no correlation 
between normal or abnormal rotation curves and the velocity dispersion 
of the group. In order to confirm this, we investigate correlations 
between dynamics and other properties of group. From previous studies 
by Hickson et al. (1988, 1992), the group size, the velocity dispersion, 
the galaxy number density, and the crossing time are listed in Table 5 
for each HCG galaxy. Figure 4(a)-(d) show 
relations between the asymmetry parameter $A$ and 
the group size (Hickson et a. 1992), the velocity dispersion 
(Hickson et al. 1992), the galaxy number density 
(Hickson et al. 1988), and the crossing time (Hickson et al. 1992) 
for case of early-type (S0/a-Sbc), late-type (Sc and later), 
``f'' rotation curves, ``fp'' ones, and ``p'' ones. 
We adopt the null hypothesis that the asymmetry parameter $A$ is not 
correlated with each group property and apply the Spearman-rank 
statistical test for all the correlations shown in Figure 4(a)-(d). 
A summary of the statistical tests is given in Table 8. 
Although the probability between $A$ and group velocity dispersion for ``f'' 
rotation curves is $4.81\times10^{-3}$, the three $\sigma$ confidence 
level is $1.3\times10^{-3}$. Therefore,   
for any cases, it is found that there is no correlation between 
the asymmetry parameter $A$ and the group properties.

\section{COMPARISON OF ROTATION CURVES BETWEEN HCG GALAXIES 
         AND FIELD GALAXIES}

\subsection{The Field Sample}

In order to investigate how the dynamical properties of HCG galaxies
are different from those of isolated galaxies, we need a reference sample
of field galaxies.
Rubin et al. (1980, 1982, 1985) published rotation curves for
60 spiral galaxies (16 Sa, 23 Sb, and 21 Sc galaxies).
Excluding two Virgo spiral galaxies (NGC 4321 and NGC 4419) and 
two galaxies in groups (NGC 1353 and NGC 4448), 
we adopt 56 spiral galaxies (15 Sa, 21 Sb, and 20 Sc) 
as a field galaxy sample. 
The basic data of these field galaxies are summarized in Table 9. 

Unfortunately, Rubin et al. (1980, 1982, 1985) show only average rotation
velocity $V(|r|)$ which is the mean of velocities within the radial bins
of both $r$ and $-r$, and 1 $\sigma$ error of the mean, $dV(|r|)$.
It is hard to know how many data points either $r$ or $-r$ contains
and to estimate $V(r)$ and $V(-r)$. 
Therefore, we simply assume that there is only one velocity data
point in each bin of $r$ and $-r$.
Making $V(r)$ and $V(-r)$ that here we reconstruct with
$V(|r|)$ and $dV(|r|)$ of Rubin et al. (1980, 1982, 1985)
under such assumption, one finds that
the rotation velocity at distance $r$ is $V(r)=V(|r|)+dV(|r|)/\sqrt{2}$
and that
the rotation velocity at distance $-r$ is $V(-r)=V(|r|)-dV(|r|)/\sqrt{2}$.
In this case we can define the asymmetry parameter $A$ as,
\begin{equation}
A \equiv \frac{1}{\sqrt{2}}\left[ \frac{1}{N}\sum^{N}_{j} \left( 
\frac{dV(|r_{j}|)}{V(|r_{j}|)} 
\right)^{2} \right]^{1/2}.  
\end{equation}
The meaning of $j$ and $N$ are the same as those for the equation (1). 
Note that the asymmetry parameters $A$ for field spirals are all 
upper limits according to the above definition. 

We also classify the shape of rotation curves for field spiral 
galaxies. The asymmetry parameter $A$ and the shape of 
the rotation curves for the field spiral galaxies are listed 
in Table 9. 

\subsection{Comparison of the Rotation Velocity between 
the HCG Spiral Galaxies and the Field Spiral Galaxies}

In Figure 5, we show the rotation curves of our HCGs spiral galaxies 
as a function of distance from galactic nucleus in units of 
$R_{25}$. We also show expected rotation curves of the field galaxies 
calculated from synthetic rotation curves in Rubin et al. (1985).
The rotation curves of the HCG spiral galaxies are different from 
those of the field spiral galaxies. Many spiral galaxies in HCGs have 
peculiar rotation curves. 

By numerical simulations, Barton et al. (1999) 
found that galaxy collision caused rising rotation curve. 
Some HCG spirals have such a rising rotation curve (e.g. HCG 61c 
and 88a). 
There are also spiral galaxies with higher rotation velocities 
(e.g. HCG 40c) and those with lower rotation velocities (e.g. HCG 88b 
and 88d) with respect to the field spirals with both the same Hubble 
type and the same luminosity. 

Rubin et al. (1991) show that the HCG spirals tend to have lower 
rotation velocity with respect to the field ones. For some HCG spirals, 
we found the same tendency. However, it is difficult to determine 
intrinsic inclination angles {\it i} of HCG spirals showing peculiar 
morphologies. Although we estimated the inclination angles {\it i} of 
the HCG spirals, there is an uncertainty of {\it i}. Most of 
the rotation curves of HCG spirals are too peculiar to discuss 
the mass distribution in the galaxies. 

\subsection{Comparison of the Asymmetry Parameter between
            the HCG Spiral Galaxies and the Field Spiral Galaxies}

We compare the frequency distributions of $A$ between the HCGs and 
the field spiral galaxies in Figure 6. 
The comparison is carried out for the case of all Hubble type, 
S0/a $-$ Sbc type, 
and Sc and later type galaxies. We include two field spiral 
galaxies with $A = 0$ into the smallest bin (NGC 2608 and IC 724). 
We find that $A$ values of the HCG spirals are 
generally larger than those of the field spirals for all 
the morphological types. 
In particular, we mention that 
the maximum value of $A$ of the field spirals 
is 0.06 (NGC 3054 and NGC 7171) while 18 HCGs spirals (53\%) have $A > 1.0$.

We apply the KS test. 
The null hypothesis is that the observed distributions of $A$
of both the HCG galaxies and the field ones
come from the same underlying population.
We obtain KS possibilities (see Table 10); 
$1.15 \times 10^{-11}$ for all the galaxies, $1.76 \times 10^{-6}$ 
for S0/a -- Sbc galaxies, and $7.40 \times 10^{-7}$ for Sc and later-type 
spiral galaxies. Therefore, we conclude that the HCG galaxies
tend to have asymmetrical rotation curves with respect to the field
galaxies. This is consistent with the result of Rubin et al. (1991).

For the HCGs spirals, we find that the late-type (Sc and later) spirals 
have larger $A$ than the early-type (S0/a $-$ Sbc) ones. 
Applying the KS test, we obtain a probability of $2.32 \times 10^{-3}$ 
between early-type HCG spirals and late-type HCG spirals 
while 0.792 for a comparison for that between early-type spirals 
in the field and late-type spirals in the field (see Table 10). 
Therefore, it is suggested that the HCG late-type spirals tend to have
more asymmetrical rotation curves with respect to 
the HCG early-type ones.

\subsection{Comparison of the Rotation Curve Shape between
            the HCG Spiral Galaxies and the Field Spiral Galaxies}

We compare the rotation curve shapes between the HCG galaxies 
and the field ones. In Figure 7 we compare the frequency 
distributions of rotation curve shapes between the HCGs spirals 
and the field ones for all galaxies, early-type spirals, 
and late-type spirals. We adopt the null hypothesis that 
the distribution for the HCGs spirals is the same as that for 
the field ones and apply the $\chi^{2}$ test. 
The results are given in Table 11. We find that there are significant 
statistical differences between the HCGs spirals and the field ones. 

Among the 39 HCG galaxies, 33 ($\approx$85\%) galaxies have either 
``fp'' or ``p'' rotation curves. 
It is remarkable that almost all the HCG late-type spirals
($\approx$93\%) have ``p'' rotation curves.
On the contrary, only 42\% of the HCG early-type spirals 
have ``p'' rotation curves and 21\%  of the HCG early-type 
spirals have ``f'' rotation curves.

The probability of the $\chi^{2}$ test between the 
early-type field spirals and late-type field spirals is 0.673, 
while that between the early-type HCG spirals and the late-type 
HCG spirals is $4.44\times10^{-3}$. 
Although the latter probability suggests a marginally significant 
difference, a larger sample will be necessary to confirm it.

\section{COMPARISON OF ROTATION CURVES BETWEEN HCG GALAXIES 
         AND CLUSTER GALAXIES}

\subsection{The Cluster Sample}

As we previously mentioned, the local galaxy number density of HCGs 
is comparable to that of cluster centers. Therefore it is interesting 
to compare the dynamical properties of the HCG spiral galaxies with 
those of cluster spiral galaxies. We use a sample of cluster spiral 
galaxies taken from Amram et al. (1992, 1994, and 1995) who published 
rotation curves of 41 spiral galaxies in eight clusters. 
The basic data of these cluster galaxies are summarized in Table 12. 
Their morphological types are taken from RC3 if available (32 galaxies). 
For the remaining nine spiral galaxies without RC3, we adopt 
the morphological types taken from Bell \& Whitmore (1989), 
Amram et al. (1995), and Gavazzi \& Boselli (1996).   

Using the equation (1), we calculated the asymmetry parameter $A$ 
of the cluster spiral galaxies. We also classified the shape of 
the rotation curves for them. The results are also listed in Table 12. 

\subsection{Comparison of the Asymmetry Parameter between 
            the HCG Spiral Galaxies and the Cluster Spiral Galaxies}

In Figure 6, we compare the frequency distribution of the asymmetry 
parameter $A$ between the HCG spiral galaxies and 
the cluster ones. 
The comparison is carried out for the case of all Hubble types, 
S0/a -- Sbc type, and Sc and later type galaxies. 
The null hypothesis is that the observed distributions of $A$ of 
both HCG spirals and the cluster ones come from the same underlying 
population. Applying the KS test, we obtain the following 
KS probabilities (see Table 13); $6.64 \times 10^{-4}$ for all 
the galaxies, 
$4.50 \times 10^{-2}$ for the S0/a -- Sbc spiral galaxies, 
and $8.48 \times 10^{-4}$ for the Sc and later-type spiral galaxies.
We find that the $A$ values of HCG spiral galaxies tend to 
be larger than those of the cluster ones for all the case. 

\subsection{Comparison of the Rotation Curve Shape between 
            the HCG Spiral Galaxies and the Cluster Spiral Galaxies}

We show the frequency distribution of the rotation curve shape 
for the HCG, the field, and the cluster spirals in Figure 7. 
We compare the both 
distributions for the case of all Hubble type, S0/a-Sbc type, 
and Sc and later-type spirals. The null hypothesis is that 
the observed distributions of the rotation curve shape of 
both the HCG spirals and the cluster ones come from the same underlying 
population. Applying the $\chi^{2}$ test, 
we obtain the following probabilities (see Table 14); $9.28 \times 10^{-7}$ 
for all the galaxies, $3.84 \times 10^{-3}$ for the S0/a-Sbc spirals, 
and $3.23 \times 10^{-5}$ for the Sc and later-type spirals. 
These results suggest that the HCG spirals tend to show more 
peculiar shapes of rotation curve.
  
\section{DISCUSSION}

In this paper, we have investigated the 
asymmetry and the shape of 
the rotation curves of spiral galaxies in HCGs.
\ \\
  
First, we investigated the relation between the morphological 
peculiarity and dynamical peculiarity of HCG spiral galaxies. 
We found that there is no statistical difference between them 
(Table 6), being consistent with the finding by Rubin et al. (1991). 
It is likely that the HCG spirals with both the morphological peculiarity 
and the dynamical peculiarity have experienced recent galaxy 
collisions. However, there are normal-morphology galaxies with peculiar 
dynamical properties and peculiar-morphology ones with normal dynamical 
properties. 
No correlation between the dynamical peculiarity and 
the morphological peculiarity suggests that the dynamical 
properties of the HCG spirals may be governed by galaxy collision 
parameters such as the difference of masses, and the orbital 
parameters.
While the morphological peculiarities such as tidal tails and 
tidal bridges, and asymmetry come to be more clearly seen 
in outer regions of galaxies, the dynamical peculiarities  
probed by the H$\alpha$ line emission are observed in inner 
regions ($r < 0.5 R_{25}$) of galaxies (see Figure 5). 
The morphological peculiarity can be more easily induced 
by galaxy collisions than the dynamical peculiarity.  
Thus, weak galaxy collisions could not perturb the galaxy rotation 
curves, although it is experienced that morphological deformation 
could be induced in outer parts of the galaxies. 
On the other hand, minor mergers could perturb the rotation curve 
in inner regions of galaxies without 
causing global morphological peculiarities. 
\ \\

Second, we investigated the relation between the properties of 
the rotation curves and the properties of 
the nuclear activity. Since it has often been suggested 
that the galaxy collisions 
trigger nuclear activities such as AGN and nuclear 
starburst phenomena (Kennicutt \& Keel 1984; 
Keel 1996), it is expected that such nuclear activities are 
more often observed in HCG galaxies. 
However, we found that there is no significantly statistical difference 
in the distribution of the asymmetry parameter $A$ between the nuclear 
activities (Figure 2). We also found that there is no correlation between 
the asymmetry parameter $A$ and $[$N{\sc ii}$]$/H$\alpha$ ratio 
(Figure 3). All these indicate that galaxy collisions do not always 
trigger the nuclear activity

Third, we investigated 
the relation between the dynamical properties of HCG spiral galaxies 
and the properties of the compact groups such as the group size, 
the velocity dispersion of member galaxies, the galaxy number density, 
and the crossing time (see Figure 4a-4d and Table 8). 
In compact groups with higher number densities and 
with smaller sizes, more frequent galaxy collision would occur and 
thus spiral galaxies in such groups would show more asymmetric 
rotation curves. 
However, we found that there are no significant correlations 
between the asymmetry parameter $A$ and any group properties.
There may be two alternative interpretations for this finding. 
1) Many compact groups containing spiral galaxies are false groups; 
e.g., a pair of galaxies with a few field galaxies 
(Mamon 1986, 1992, 1994, 1995; Hernquist et al. 1995). 
2) The projected size and the observed velocity dispersion 
of the compact groups are significantly different from 
the real (i.e., three-dimensional) ones. 
As mentioned in section 5.4, 85\% of HCG spirals are peculiar and 
that 93 \% of the late type HCG spirals are peculiar. This high rate 
of peculiarity seems inconsistent with the false group hypothesis. 
However, these two possibilities 
will be taken into account in future studies. 

Finally, we compared the dynamical properties of spiral galaxies 
among the field, the clusters, and the HCGs. We found that the HCG spirals 
tend to have more asymmetric and more peculiar rotation curves 
with respect to the field and the cluster spirals 
(see Figure 6 and 7, Table 10, 11, 13, and 14). 
It is interesting to note that these are the significant differences 
in the distributions of the $A$ and the rotation curve morphologies 
between the HCG early-type spirals and the HCG late-type ones, 
although we found no such difference between 
the field early-type spirals and the field late-type ones.  
This may be attributed to the stabilization 
of a disk by a large bulge of early-type spirals
(Mihos \& Hernquist 1994; Vel\'{a}zquez \& White 1999). 
Since a small bulge cannot stabilize the disk, 
late-type spirals are more sensitive to galaxy 
interactions than the early-type spirals and are expected to show 
the peculiar kinematical properties for a longer duration. 

\begin{acknowledgements}
We are grateful to the staff of OAO for kind help of the observations.
We would like to thank Daisuke Kawata and Tohru Nagao for useful comments 
and suggestions. 
TM is thankful for support from a Research Fellowship from the Japan
Society for the Promotion of Science for Young Scientists.
This work was partly supported by the Ministry of Education, Science, 
Culture, and Sports (Nos. 07044054, 10044052, and 10304013).
\end{acknowledgements}


\newpage

\figcaption{
Major-axis velocities with respect to the systemic 
heliocentric velocity for 30 HCG galaxies as a function of projected 
nuclear distance. 
\label{fig1}
}

\figcaption{
Comparisons of frequency distributions of the logarithmic asymmetric
parameter (log $A$) between the AGN (upper panels) 
and H{\sc ii} nuclei (lower panels).
\label{fig2}
}

\figcaption{
The relations between the logarithmic asymmetry parameter $A$ 
and the $[$N{\sc ii}$]$/H$\alpha$ ratio. The filled circles 
and open circles indicate the early-type (S0/a-Sbc) and 
the late-type (Sc and later) HCG spiral galaxies. The open squares, 
filled squares, and crosses indicate 
the ``f'', ``fp'', and ``p'' rotation curves, respectively. 
\label{fig3}
}

\figcaption{
The relations between the logarithmic asymmetry parameter $A$ 
and a) the size, b) the velocity dispersion, c) galaxy number 
density, and d) crossing time of group. The filled circles 
and open circles indicate the early-type (S0/a-Sbc) and 
the late-type (Sc and later) HCG spiral galaxies. The open squares, 
filled squares, and crosses indicate 
the ``f'', ``fp'', and ``p'' rotation curves, respectively. 
\label{fig4}
}

\figcaption{
Major-axis velocities with respect to the systemic 
heliocentric velocity for 30 HCG galaxies as a function of 
nuclear distance in the unit of $R_{25}$. The open circles and filled 
circles indicate approaching side and going away side. 
The middle line of three indicates expected rotation curves of 
the field spiral galaxies with similar Hubble type and luminosity 
calculated from synthetic rotation curves in Rubin et al. (1985).
The top and bottom line of three are expected rotation curves 
of the field spiral galaxies with similar Hubble type and one magnitude 
brighter and fainter luminosity.    
\label{fig5}
}

\figcaption{
Comparisons of frequency distributions of the logarithmic asymmetric
parameter (log $A$) between the HCG spirals (upper three panels), 
the field ones (middle three panels), and the cluster spirals 
(lower three panels), for all Hubble type, early-type (S0/a-Sbc), 
and late-type (Sc and later).
\label{fig6}
}

\figcaption{
Comparison of frequency distributions of the rotation curve shapes
between the HCG spirals (upper panel), the filed ones (middle panel), 
and the cluster ones (lower panel) for all Hubble type, early-type 
(S0/a-Sbc), and late-type (Sc and later). 
\label{fig7}
}

\appendix

\section{Velocities for spirals in Hickson compact groups}

Table 15 lists the distances from galactic nucleus $r$ in units of 
arcsec and of $R_{25}$, the rotation velocities $V(r)$ at $r$, and 
their 1 $\sigma$ errors $dV(r)$ for each observed HCG spirals.
\end{document}


%
\begin{deluxetable}{ccrr}
\tablecaption{A journal of observations}
\tablehead{
\colhead{}    & \colhead{}     & \colhead{Exp.}  & \colhead{PA} \nl
\colhead{HCG} & \colhead{Date} & \colhead{(sec)} & \colhead{(deg)}
}
\startdata
7a  & 19 Aug 1996 & 1800 & 159 \nl
31a & 09 Jan 1997 & 1800 &  80 \nl
31b & 09 Jan 1997 & 1800 &  41 \nl
31c & 09 Jan 1997 &  860 &  41 \nl
37b & 21 Feb 1996 & 1800 &  77 \nl
38a & 06 Jan 1997 & 1800 & 340 \nl
40c & 23 Feb 1996 & 1800 & 122 \nl
44a & 21 Feb 1996 & 1800 & 125 \nl
44c & 21 Feb 1996 & 1800 & 130 \nl
44d & 21 Feb 1996 & 1800 & 125 \nl
47a & 21 Feb 1996 & 1800 &  15 \nl
53a & 20 Feb 1996 & 1800 &  93 \nl
61c & 21 Feb 1996 & 1800 & 130 \nl
68c & 21 Feb 1996 & 1800 &  18 \nl
71a & 20 Feb 1996 & 1800 &  23 \nl
73a\tablenotemark{a} 
    & 25 Feb 1996 & 1800 &  18 \nl
79d & 18 Aug 1996 & 1800 &   0 \nl
80a & 25 Feb 1996 & 1800 &  64 \nl
87a\tablenotemark{b} 
    & 18 Aug 1996 & 1800 &  56 \nl
87c & 19 Aug 1996 & 1800 &  90 \nl
88a & 15 Aug 1996 & 1800 & 132 \nl
88b & 15 Aug 1996 & 1800 &  31 \nl
88c & 15 Aug 1996 & 1800 &  31 \nl
88d & 18 Aug 1996 & 1800 &  71 \nl
89a & 19 Aug 1996 & 1800 &  57 \nl
92a\tablenotemark{a} 
    & 15 Aug 1996 & 1800 & 137 \nl
92c & 18 Aug 1996 & 1800 & 140 \nl
93b & 19 Aug 1996 & 1800 & 143 \nl
93c & 19 Aug 1996 & 1800 &  98 \nl
96a & 19 Aug 1996 &  480 &  90 \nl
\enddata
\tablenotetext{a}{HCG 73a and 92a are redshift discordant galaxies, 
and we exclude these galaxies from statistical sample.}
\tablenotetext{b}{HCG 87a is classified into lenticuler galaxy in RC3, 
and we exclude this galaxy from statistical sample.}
\end{deluxetable}

\clearpage
%
\begin{deluxetable}{ccc}
\tablecaption{The Asymmetry parameters and the shape parameter 
              of the rotation curves for the observed galaxies}
\tablehead{
\colhead{HCG} & \colhead{$A$} & \colhead{shape}
}
\startdata
7a  & 0.23 & p  \nl
31a & 0.12 & p  \nl
31b & 0.27 & p  \nl
31c & 0.38 & p  \nl
37b &      & ?  \nl  
38a & 0.07 & fp \nl
40c & 0.09 & f  \nl
44a & 0.09 & p  \nl  
44c & 0.13 & fp \nl
44d & 6.55 & p  \nl
47a &      & p  \nl  
53a & 0.06 & f  \nl
61c &      & fp \nl  
68c & 0.11 & fp \nl
71a & 6.23 & p  \nl
73a & 0.18 & p  \nl          
79d & 0.84 & p  \nl
80a & 0.40 & p  \nl
87a &      & p  \nl          
87c & 0.26 & p  \nl
88a & 0.26 & fp \nl
88b & 0.49 & p  \nl
88c & 1.21 & p  \nl
88d & 0.15 & p  \nl
89a & 0.14 & fp \nl
92a & 0.41 & p  \nl
92c &      & ?  \nl  
93b & 0.32 & p  \nl
93c & 0.34 & p  \nl
96a &      & p  \nl  
\enddata
\end{deluxetable}

\clearpage
%
\begin{deluxetable}{cccc}
\tablecaption{Comparison of kinematical properties with morphological peculiarities}
\tablehead{
\colhead{}           & \multicolumn{3}{c}{Rotation Curves} \nl
\colhead{Morphology} & \colhead{f} & \colhead{fp} & \colhead{p} \nl
}
\startdata
normal    & 2 & 2 &  8 \nl
peculiar  & 4 & 7 & 16 \nl
\hline
P($\chi^{2}$)\tablenotemark{a}
          & \multicolumn{3}{c}{0.818}
\enddata
\tablenotetext{a}{The possibility that rejects the null hypothesis.}
\end{deluxetable}

\clearpage
\begin{deluxetable}{ccc}
\tablecaption{A summary of the Spearman rank test for 
               the correlations between asymmetry parameter $A$ 
                 and the $[$N{\sc ii}$]$/H$\alpha$ ratio}
\tablehead{
\colhead{} & \colhead{Number} 
           & \colhead{Probabilities\tablenotemark{a}}
}
\startdata
All      & 20 & 0.243 \nl
\hline
S0/a--Sbc & 10 & 0.987 \nl
Sc$\leq$ & 10 & 0.907 \nl
f        &  3 & 0.667 \nl
fp       &  4 & 0.200 \nl
p        & 13 & 0.566 \nl
\enddata
\tablenotetext{a}{The probability that rejects the null hypothesis.}
\end{deluxetable}

\clearpage
\begin{deluxetable}{cccccc}
\tablecaption{A summary of the Spearman rank test for the correlations
               between asymmetry parameter $A$ and the properties 
                 of the HCGs}
\tablehead{
\colhead{}       & \colhead{} 
                 & \multicolumn{4}{c}{Probabilities\tablenotemark{a}} \nl
\colhead{}       & \colhead{Number} & \colhead{$\log R$} 
                 & \colhead{$\log \sigma_{\rm r}$}
                 & \colhead{$\log \rho_{\rm N}$}
                 & \colhead{$\log H_{0}t_{\rm c}$}
}

\startdata
All       & 34 & 0.737 & 0.355               & 0.748 & 0.804 \nl
\hline
S0/a--Sbc & 20 & 0.686 & 0.052               & 0.135 & 0.028 \nl
Sc$\leq$  & 14 & 0.543 & 0.037               & 0.533 & 0.085 \nl
f         &  6 & 0.329 & $4.81\times10^{-3}$ & 0.397 & 0.266 \nl
fp        &  8 & 0.955 & 1.000               & 0.867 & 1.000 \nl
p         & 20 & 0.701 & 0.519               & 0.762 & 0.565 \nl
\enddata
\tablenotetext{a}{The probability that rejects the null hypothesis.}
\end{deluxetable}

\clearpage
%
\begin{deluxetable}{cccccccc}
\tablecaption{Results of KS test for distributions 
              of the asymmetry parameter between 
              the HCG spiral galaxies and the field spiral galaxies }
\tablehead{
\colhead{} & \colhead{KS Probability\tablenotemark{a}} 
}
\startdata
All HCG        vs. All Field      & $1.15\times10^{-11}$ \nl
HCG S0/a-Sbc   vs. Field S0/a-Sbc & $1.76\times10^{-6}$  \nl
HCG Sc-        vs. Field Sc-      & $7.40\times10^{-7}$  \nl
HCG S0/a-Sbc   vs. HCG Sc-        & $2.32\times10^{-3}$  \nl
Field S0/a-Sbc vs. Field Sc-      & $0.792$              \nl
\enddata
\tablenotetext{a}{The possibility that rejects the null hypothesis.}
\end{deluxetable}

\clearpage
%
\begin{deluxetable}{cccccccc}
\tablecaption{Frequency distributions of the rotation curve shape  
              between the HCGs spiral galaxies 
              and the field spiral galaxies\tablenotemark{a}}
\tablehead
{
    &
    \multicolumn{2}{c}{All\tablenotemark{b}} &
    \multicolumn{2}{c}{S0a-Sbc}              &
    \multicolumn{2}{c}{Sc-}                  \nl
    &
    \colhead{HCGs}  &
    \colhead{Field} &
    \colhead{HCGs}  &
    \colhead{Field} &
    \colhead{HCGs}  &
    \colhead{Field} &
}
\startdata
f  & 15.4  (6) & 69.1 (38) & 20.8  (5) & 65.8 (25) &  6.7  (1) & 76.5 (13) \nl
fp & 23.1  (9) & 25.5 (14) & 37.5  (9) & 28.9 (11) &  0.0  (0) & 17.6  (3) \nl
p  & 61.5 (24) &  5.5  (3) & 41.7 (10) &  5.3  (2) & 93.3 (14) &  5.9  (1) \nl
\hline
P($\chi^{2}$)\tablenotemark{c} & 
\multicolumn{2}{c}{$3.23 \times 10^{-9}$} &
\multicolumn{2}{c}{$2.55 \times 10^{-4}$} & 
\multicolumn{2}{c}{$4.73 \times 10^{-6}$} 
\enddata
\tablenotetext{a}{Numbers in parentheses are the actual numbers.}
\tablenotetext{b}{``All'' means that the total sample of S0a$-$Sbc, 
                  and Sc- galaxies .}
\tablenotetext{c}{The possibility that rejects the null hypothesis.}
\end{deluxetable}

\clearpage
%
\begin{deluxetable}{cccccccc}
\tablecaption{Results of KS test for distributions 
              of the asymmetry parameter between 
              the HCG spiral galaxies and the cluster spiral galaxies }
\tablehead{
\colhead{} & \colhead{KS Probability\tablenotemark{a}} 
}
\startdata
All HCG        vs. All Cluster      & $6.64\times10^{-4}$ \nl
HCG S0/a-Sbc   vs. Cluster S0/a-Sbc & $4.50\times10^{-2}$  \nl
HCG Sc-        vs. Cluster Sc-      & $8.48\times10^{-4}$  \nl
HCG S0/a-Sbc   vs. HCG Sc-          & $2.32\times10^{-3}$  \nl
Cluster S0/a-Sbc vs. Cluster Sc-    & $0.183$              \nl
\enddata
\tablenotetext{a}{The possibility that rejects the null hypothesis.}
\end{deluxetable}

\clearpage
%
\begin{deluxetable}{cccccccc}
\tablecaption{Frequency distributions of the rotation curve shape  
              between the HCGs spiral galaxies 
              and the clusters spiral galaxies\tablenotemark{a}}
\tablehead
{
    &
    \multicolumn{2}{c}{All\tablenotemark{b}} &
    \multicolumn{2}{c}{S0a-Sbc}              &
    \multicolumn{2}{c}{Sc-}                  \nl
    &
    \colhead{HCGs}  &
    \colhead{Clusters} &
    \colhead{HCGs}  &
    \colhead{Clusters} &
    \colhead{HCGs}  &
    \colhead{Clusters} &
}
\startdata
f  & 15.4  (6) & 70.7 (29) & 20.8  (5) & 64.3 (18) &  6.7  (1) & 84.6 (11) \nl
fp & 23.1  (9) & 17.1  (7) & 37.5  (9) & 25.0  (7) &  0.0  (0) &  0.0  (0) \nl
p  & 61.5 (24) & 12.2  (5) & 41.7 (10) & 10.7  (3) & 93.3 (14) & 15.4  (2) \nl
\hline
P($\chi^{2}$)\tablenotemark{c} & 
\multicolumn{2}{c}{$9.28 \times 10^{-7}$} &
\multicolumn{2}{c}{$3.84 \times 10^{-3}$} & 
\multicolumn{2}{c}{$3.23 \times 10^{-5}$} 
\enddata
\tablenotetext{a}{Numbers in parentheses are the actual numbers.}
\tablenotetext{b}{``All'' means that the total sample of S0a$-$Sbc, 
                  and Sc- galaxies .}
\tablenotetext{c}{The possibility that rejects the null hypothesis.}
\end{deluxetable}%
